\documentstyle[12pt]{article}
\def\MeV{\,{\rm MeV}}
\def\sec{\,{\rm sec}}

\def\eV{{\,\rm eV}}
\def\erg{{\,\rm erg}}
\def\cmm2{{\,\rm cm^{-2}}}
\def\cm2{{\,{\rm cm}^2}}
\def\cmm3{{\,{\rm cm}^{-3}}}
\def\gcmm3{{\,{\rm g\,cm^{-3}}}}

\def\VEV#1{\left\langle #1\right\rangle}
\def\GF{G_{\rm F}}

\def\rns{r_{\rm ns}}
\def\P{{\bf P}}

\def\PLZ{P_{\rm LZ}}
\def\Neff{N_{\rm eff}}
\def\Er{E_{\rm r}}
\def\Nod{N_{\rm od}}
\def\Geff{\gamma_{\rm eff}}
\def\umin{u_{\rm min}}
\def\umax{u_{\rm max}}
\def\PE{{\bf P}_E}
\def\VE{{\bf V}_E}
\def\la{\mathrel{\mathpalette\fun <}}
\def\ga{\mathrel{\mathpalette\fun >}}
\def\fun#1#2{\lower3.6pt\vbox{\baselineskip0pt\lineskip.9pt
  \ialign{$\mathsurround=0pt#1\hfil##\hfil$\crcr#2\crcr\sim\crcr}}}
\begin{document}
\pagestyle{empty}
\begin{center}
\vspace*{3cm}
\rightline{FERMILAB--Pub--94/369-A}

\vspace{.2in}
{\Large \bf NEUTRINO MIXING CONSTRAINTS \\
\bigskip
AND SUPERNOVA NUCLEOSYNTHESIS} \\

\vspace{.2in}
G.~Sigl$^{1,2}$\\
\vspace{.2in}
{\it $^1$Department of Astronomy \& Astrophysics\\
Enrico Fermi Institute, The University of Chicago, Chicago, IL~~60637-1433}\\

\vspace{0.1in}
{\it $^2$NASA/Fermilab Astrophysics Center\\
Fermi National Accelerator Laboratory, Batavia, IL~~60510-0500}\\

\end{center}

\vspace{.3in}

\centerline{\bf ABSTRACT}
\medskip
We reexamine the constraints on mixing between electron and muon or tau
neutrinos from shock-reheating and r-process nucleosynthesis in supernovae.
To this end neutrino flavor evolution is described by nonlinear equations
following from a quantum kinetic approach. This takes into account neutrino
forward scattering off the neutrino background itself. In contrast to other
claims in the literature it is shown that a sound self-consistent analytical
approximation can in the first place only be performed in the adiabatic limit
where phase terms are suppressed. In the cosmologically interesting mass range
below about $25\eV$ the resulting mixing parameter bounds are between one and
two orders of magnitude less restrictive than limits neglecting neutrino
contributions. Extensions to the non-adiabatic regime derived in the
literature usually neglect coherence effects from phases. To check their
importance numerical simulations of the evolution equations were performed in
this regime. They indicate that analytical approximations for the flavor
conversion efficiencies can indeed be extended by neglecting phase terms. This
allows more stringent bounds similar to the ones derived in earlier work.
These bounds depend to some extent on the adopted supernova model and tend to
be somewhat less restrictive in the mixing angle but simultaneously extend to
smaller mixing masses compared to limits neglecting the neutrino induced
potential.
\newpage
\pagestyle{plain}
\setcounter{page}{1}
\newpage

\section{Introduction}
Neutrino oscillations became interesting in astrophysics not
least because of the MSW effect~\cite{Mikheyev} which due to a
cancellation of a small vacuum mixing term and a
flavor dependent forward scattering amplitude can lead
to medium enhanced conversion between different neutrino flavors.
For suitable mixing parameters this effect offers a solution to
the solar neutrino problem~\cite{Bethe,Haxton,Parke,Kuo1}.
Medium enhanced neutrino oscillations were
also discussed in the circumstance of supernova
explosions~\cite{Fuller,Kuo2,Notzold,Qian}. As long as the neutrino
densities are much smaller than the electron density, as in the
sun and in the outer envelope of a supernova, neutrino forward
scattering off the neutrino background itself can safely be
neglected in the MSW analysis. In the opposite extreme case,
however, it has been demonstrated for the case of neutrino mixing
in the early universe that interactions among the neutrinos can
change the character of the oscillations
drastically~\cite{early}.

There has been a discussion~\cite{Panta,Qian1,Cline} whether similar
effects could be important for oscillations among $\nu_e$ and
$\nu_\mu$ or $\nu_\tau$ neutrinos
in the hot bubble region above the neutrinosphere after
supernova core bounce. This is especially important since
allowing r-process nucleosynthesis in this region to work a few
seconds after core bounce forbids possible resonance transitions
of the more energetic $\nu_\mu$ or $\nu_\tau$ neutrinos into
$\nu_e$ neutrinos to occur with high efficiency~\cite{Qian}. It is
one of the rare astrophysical situations where neutrino masses
relevant for cosmological hot dark matter
candidates, between $1\eV$ and $100\eV$, play a role. Furthermore it has been
suggested~\cite{Fuller} that oscillations between the
neutrinosphere and the outward going shock could have an important
impact on the delayed shock heating mechanism itself. Ref.~\cite{Qian1}
contains an extensive discussion of the bounds on neutrino mixing
which can be derived by considering these two situations.
However, their approach suffers in part from the presence of
ambiguous oscillation phases which are hard to implement analytically.
We therefore found it worth to reexamine this problem
by a combination of analytical and numerical work.

In section 2 we set up the flavor evolution equations for this
problem as they follow from a quantum kinetic description of
neutrino oscillations. In section 3 we describe the physical
situation in the postbounce supernova during the parts of the
cooling phase which are relevant to us. Section 4 reexamines the
case of highly adiabatic neutrino oscillations where
ambiguous oscillation phases are suppressed allowing an
analytical approach for the transition efficiency. Reliability
and model dependence of the resulting bounds on the mixing parameters
are discussed.
In section 5 we perform numerical simulations and show that
analytical estimates for flavor conversion efficiencies can
still be used to extend these bounds to the nonadiabatic regime.
We summarize our results in section 6.

\section{Quantum Kinetic Flavor Evolution}
As was shown in Ref.~\cite{Sigl} an ensemble of neutrinos and
antineutrinos consisting of $N$ relativistic mixed
flavors can be described by a set of $N\times N$ density
matrices $\rho_{\bf p}$ and $\bar\rho_{\bf p}$ (overbarred
quantities refer to antineutrinos from now on), one for each
momentum mode ${\bf p}$.
The $i$th diagonal term represents the occupation number for
neutrinos of flavor $i$ in this mode whereas the off diagonal
terms describe the coherence of the mixing flavors. The
evolution equations for the $\rho_{\bf p}$ were given in
Ref.~\cite{Sigl} in the most general case. For the problem under
consideration here various simplifications can be applied.

First, we
restrict ourselves to two flavor mixing between the electron
neutrino and the muon or tau neutrino. The transformation
between the flavor eigenstates $\nu_e$ and $\nu_\mu$ on the one
hand and the mass eigenstates $\nu_1$ and $\nu_2$ on the other
hand is then characterized by the vacuum mixing angle $\theta$
via
\begin{equation}
  \left({\nu_e\atop\nu_\mu}\right)=\left(\begin{array}{cc}
  \cos\theta & -\sin\theta \\
  \sin\theta & \cos\theta \end{array}\right)
  \left({\nu_1\atop\nu_2}\right)\,,\label{trafo}
\end{equation}
where we adopt the convention that $\theta<\pi/4$. Matter
induced resonances can lead to proton rich conditions (i.e. the
number of electrons per baryon $Y_e$ is bigger than 0.5) in r-process
nucleosynthesis if they occur for neutrinos but not for
antineutrinos~\cite{Qian}. In the above convention this is the case if
$\Delta=m^2_2-m^2_1>0$ where $m_{1,2}$ are the mass eigenvalues
corresponding to $\nu_{1,2}$.

Furthermore, for $\Delta\leq10^4\eV^2$ the
resonances occur well {\it above} the neutrinosphere where
nonforward scattering and thus oscillation damping~\cite{RS} is
negligible. Thus we
can restrict ourselves to the coherent effects caused by
forward scattering off of electrons, nucleons, nuclei and the
neutrinos themselves.

In order to write the evolution equations in a convenient way we
write the density matrices
$\rho_{\bf p}$ in terms of polarizations ${\bf P}_{\bf p}$ and
Pauli matrices $\tau$
\begin{equation}
  \rho_{\bf p}={1\over2}n_{\bf p}
  \left(1+{\bf P}_{\bf p}\cdot\tau\right)\,,\label{rhop}
\end{equation}
where $n_{\bf p}={\rm Tr}(\rho_{\bf p})$. An analogous definition
holds for antineutrinos. Since for the mixing parameters under
consideration the oscillation length is always short compared to
the scale height of the neutrino density we can write down the
evolution equations for the ${\bf P}_{\bf p}$ in the
form~\cite{quakin}
\begin{equation}
  {d\over ds}{\bf P}_{\bf p}=\left[\left(\begin{array}{c}
  -{\Delta\over2\vert{\bf p}\vert}\sin2\theta \\ 0 \\
  \sqrt2\GF N_e-{\Delta\over2\vert{\bf p}
  \vert}\cos2\theta \end{array}\right)+
  \sqrt2\GF\int d{\bf q}(1-\cos\theta_{\bf pq})
  \left(n_{\bf q}{\bf P}_{\bf q}-
  {\bar n}_{\bf q}{\bar\P}_{\bf q}\right)\right]
  \times{\bf P}_{\bf p}\label{evo1}
\end{equation}
[notation $d{\bf q}\equiv d^3{\bf q}/(2\pi)^3$]. Here, $s$ is
the length measured along the path of a neutrino in mode ${\bf
p}$, $\GF$ is Fermi's constant, $N_e=N_{e^-}-N_{e^+}$ is the
difference of the
electron and positron densities $N_{e^-}$ and $N_{e^+}$
and $\theta_{\bf pq}$ is the angle
between ${\bf p}$ and ${\bf q}$. Only terms CP-odd in the
background enter the effective potential; the CP-even
contributions which can be important in the early universe are
negligible here. The term in big braces on the r.h.s. of
Eq.~(\ref{evo1}) can be regarded as the effective potential and
depends on $s$ explicitly via $N_e$ as well as implicitly
via ${\bf P}_{\bf p}$. The latter fact renders Eq.~(\ref{evo1})
a nonlinear differential equation for the polarizations
${\bf P}_{\bf p}$ in the individual modes which experience
resonances where the third component of this effective potential
vanishes.

In principle, there is also an equation for the
antineutrino flavor evolution but as we will see below even if one
takes into account the neutrino contribution to their effective
potential the antineutrinos never experience resonances.
Therefore, for small vacuum mixing angles the
fast oscillating antineutrino contribution to the effective potential
of the neutrinos can be averaged over.

For a fixed energy $E\equiv\vert{\bf p}\vert$ neutrinos from
different directions will resonate at different positions. This
tends to wash out the oscillation phases and the distribution
functions in energy space. Assuming a spherically symmetric
supernova the latter effect is negligible since the neutrinos are
almost radially free streaming. In the adiabatic limit where
phases play no role it is therefore a good approximation to
substitute $s$ by the radius $r$ and $\cos\theta_{\bf pq}$ by its
flux averaged value $F(r)$. This leads to an equation which only
depends on $E$:
\begin{equation}
  {d\over dr}\PE=\VE\times\PE=\left[\left(\begin{array}{c}
  -{\Delta\over2E}\sin2\theta \\ 0 \\
  \sqrt2\GF N_e-{\Delta\over2E}\cos2\theta
  \end{array}\right)+
  \sqrt2\GF F(r)(\P-\bar\P)\right]\times{\bf P}_E\,.\label{evo2}
\end{equation}
The self-interaction term in the effective potential reduces to
a product of the difference of the total neutrino and
antineutrino flavor polarizations $\P-\bar\P=\int d{\bf q}
(n_{\bf q}\P_{\bf q}-{\bar n}_{\bf p}{\bar\P}_{\bf p})$ and the
geometric factor $F(r)$ which for
radii $r$ large compared to the neutrinosphere radius $\rns$ is
given by~\cite{Panta}
\begin{equation}
  F(r)={1\over4}\left({\rns\over r}\right)^2\,.\label{Fr}
\end{equation}
Beyond the adiabatic regime the problem is in principle two
dimensional and the one dimensional simplification Eq.~(\ref{evo2})
tends to exaggerate the influence of coherent phase effects.
However, since we want to demonstrate that even strong coherence
effects do not change conversion efficiencies considerably we can
still use Eq.~(\ref{evo2}) as an extreme case complementary to the
approach in Ref.~\cite{Qian1} which neglected phases entirely.
Therefore, Eq.~(\ref{evo2}) will be the basic equation from which
we start our analysis below.

\section{Shock-Reheating Epoch and Hot Bubble Phase}
The energy spectrum $F_\alpha(E)$ (in units of number density
per energy) of the different neutrino species
$\alpha=\nu_e,\bar\nu_e,\nu_\mu$ assuming no
oscillations is given by numerical supernova models and is
proportional to $r^{-2}$ in the free streaming region. The spectra
of $\bar\nu_\mu$, $\nu_\tau$ and $\bar\nu_\tau$ are equal to
that of $\nu_\mu$. One generic feature is that the luminosity
$L$ is the same for all species within about $10\%$. However,
the average energies are different. The
total number densities $N_\alpha$ are then inverse proportional
to $\VEV{E_\alpha}$. On the other hand,
$F_{\nu_\mu}(E)>F_{\nu_e}(E)$ for $E\geq20\MeV$ because the muon
and tau neutrinos have higher average energy. This is the
reason why adiabatic resonance transitions between $\nu_e$ and
$\nu_\mu$ or $\nu_\tau$ would lead to more high energetic
electron neutrinos. As already mentioned in
the introduction there are two interesting
phases to distinguish during cooling of the newly born neutron
star where this effect could play an important role.

First, around $0.15\sec$ after core bounce the high neutrino
luminosity $L\sim5\times10^{52}\erg/\sec$ is expected to help
reenergize the stalled shock and lift it outward~\cite{Fuller}.
Due to charged current reactions with nucleons and nuclei the
electron neutrinos play the main role in this process. Since the
corresponding cross sections are proportional to the square of the
electron neutrino energy resonant conversion of muon or tau
neutrinos with energies $\geq25\MeV$ into electron neutrinos
could make shock revival more efficient.

R-process nucleosynthesis takes place a few seconds after core
bounce which constitutes the phase we are interested in in
this paper.
In this case resonant conversion between $\nu_e$ and $\nu_\mu$
or $\nu_\tau$ would lead to proton rich
conditions in r-process nucleosynthesis~\cite{Qian} and the
cooling supernova remnant would thus be lost as a site for this
process to occur efficiently. For treating this phase we will
use $L\sim10^{51}\erg/\sec$ as well as
Fermi-Dirac neutrino distributions with
$\VEV{E_{\nu_e}}\sim11\MeV$, $\VEV{E_{\bar\nu_e}}\sim16\MeV$ and
$\VEV{E_{\nu_\mu}}\sim25\MeV$, respectively, and vanishing
chemical potential.

\section{Analytical Approach in the Adiabatic Limit}
In this section we {\it assume} in the first place that the
transitions are adiabatic, then calculate the effective
potential from which we derive the adiabaticity index
self-consistently and finally put constraints on this index.
This is
{\it not} the same approach as adopted in Ref.~\cite{Panta}
where the neutrino induced potential was treated as a small
perturbation and results in different conclusions. A similar but
more laborious approach than ours was adopted in Ref.~\cite{Qian1}.

To begin with let us define
\begin{equation}
  N_\alpha(E)=\int_0^E dE^\prime F_\alpha(E^\prime)\,,\label{NE}
\end{equation}
and the total number density $N_\alpha=N_\alpha(\infty)$
for all neutrino species $\alpha$. In the adiabatic limit the
evolved energy dependent polarizations appearing in Eq.~(\ref{evo2})
are given by
\begin{equation}
  \P_E={F_{\nu_\mu}(E)-F_{\nu_e}(E)\over F_{\nu_\mu}(E)+
  F_{\nu_e}(E)}\left(\begin{array}{c}
  \sin2\theta_m \\ 0 \\ \cos2\theta_m \end{array}\right)
  \,,\label{Pad}
\end{equation}
with an analogous equation for antineutrinos with an overall sign
change. Here, the also energy dependent mixing angle $\theta_m$
in the medium is formally given by the components of the
effective potential $\VE$ in Eq.~(\ref{evo2}),
\begin{equation}
  \sin2\theta_m={V_{E1}\over\left(V^2_{E3}+V^2_{E1}\right)^{1/2}}
  \,,\label{formsin}
\end{equation}
and will self-consistently be determined below in
Eq.~(\ref{sinfin}). An analogous expression holds for the
antineutrino medium mixing angle ${\bar\theta}_m$.

It should be
noted that in cases different from the adiabatic limit the
column vector in Eq.~(\ref{Pad}) has to be replaced by
\begin{equation}
  \left(\begin{array}{c}
  \left(1-2\PLZ\right)\sin2\theta_m+2\left[\PLZ\left(1-\PLZ\right)
  \right]^{1/2}\cos2\theta_m\cos\alpha \\
  -2\left[\PLZ\left(1-\PLZ\right)\right]^{1/2}\sin\alpha \\
  \left(1-2\PLZ\right)\cos2\theta_m-2\left[\PLZ\left(1-\PLZ\right)
  \right]^{1/2}\sin2\theta_m\cos\alpha
  \end{array}\right)\,,\label{phases}
\end{equation}
where $\alpha$ is a phase which builds up at and beyond the
resonance and $\PLZ$ is the (Landau-Zener) transition probability
between mass
eigenstates at the resonance which therefore characterizes its
adiabaticity. Obviously, modes near resonance contribute most to
the first and second component of total polarization ${\bf
P}$ which enters into the effective potential. For $\PLZ$ not near
to 0 or 1 the phase dependent term can be important compared to the
phase independent term. The analytical treatment of this section
is therefore completely safe only in the adiabatic limit
$\PLZ\ll1$ where Eq.~(\ref{Pad}) is a good approximation. For the general
case we will resort to numerical modeling in the next section. It should
be stressed that even if for some reason the phase dependent terms should
average out (which is not clear since in the nonadiabatic case the
phases near resonance are of order unity by definition) in contrast to
the claim in Ref.~\cite{Cline}, Eq.~(\ref{phases})
still contributes a nonvanishing off diagonal component to the
effective potential in the flavor basis.

Coming back to the adiabatic case the third component of the
effective potential can be written as
\begin{equation}
  V_{E3}=\sqrt2\GF\Neff-{\Delta\over2E}\cos2\theta
  \,.\label{VE3}
\end{equation}
Here the effective density $\Neff$ including the neutrino
background itself is defined as follows: At a given radius $r$
a specific mode with energy $\Er=\Er(r)$ will be in
resonance, i.e. $V_{\Er 3}(r)=0$. Then, without knowing the
exact energy dependence of the medium mixing angle for small
vacuum mixing angles we can use
that $\theta_m\to0$ for $E<\Er$, $\theta_m\to\pi$ for $E>\Er$
and $\cos2{\bar\theta}_m\sim1$ for all energies at radius $r$. Using
this in Eq.~(\ref{evo2}) yields the approximation
\begin{equation}
  \Neff(r)=N_e+F(r)\left[N_{\nu_e}-N_{{\bar\nu}_e}+
  2N_{\nu_\mu}(\Er)-2N_{\nu_e}(\Er)\right]\,,\label{Neff}
\end{equation}
where we suppress the $r$ dependence of all number densities on
the r.h.s. Note that in this expression $\Er$ has to be considered
as a function of $r$. The first term in Eq.~(\ref{Neff}) is given by
$N_e=Y_e\rho/m_N$ ($m_N$
is the nucleon mass) with $Y_e\sim0.4$ the number of electrons
per baryon and $\rho(r)$ the density profile for which we take
the one given in Ref.~\cite{Qian} for the hot bubble phase.

In order to proceed with the off diagonal part of the effective
potential let us first formally define the ``off diagonal
density''
\begin{equation}
  \Nod(r)={F(r)\over\sin2\theta}\int_0^\infty dE\left[
  \sin2\theta_m\left(F_{\nu_\mu}(E)-F_{\nu_e}(E)\right)+
  \sin2{\bar\theta}_m\left(F_{{\bar\nu}_\mu}(E)-
  F_{{\bar\nu}_e}(E)\right)\right]\,.\label{Nod}
\end{equation}
Then using Eq.~(\ref{evo2}) we can write the first component of
the effective potential as
\begin{equation}
  V_{E1}=\left(\sqrt2\GF\Nod-{\Delta\over2E}\right)
  \sin2\theta\,.\label{VE1}
\end{equation}
We now choose a specific neutrino mode with energy $E_0$ which
resonates at $r=r_0$, i.e. $E_0=\Er(r_0)$. Then
using Eqs.~(\ref{formsin}), (\ref{VE3}) and (\ref{VE1})
allows us to write down an expression for $\theta_m$ at $r=r_0$
where $V_{E_03}(r_0)=0$:
\begin{equation}
  \sin2\theta_m={\sin2\theta\over\left[\left(
  {E/E_0-1\over2\sqrt2\GF E\Nod/\Delta-1}\right)^2
  \cos^22\theta+\sin^22\theta\right]^{1/2}}
  \,.\label{sinfin}
\end{equation}
An analogous equation with an effective sign change of $E$ holds for
${\bar\theta}_m$. Inserting Eq.~(\ref{sinfin}) into (\ref{Nod}) results
in a nonlinear equation for $\Nod$ which has to be solved after
determining the resonance point from the condition $V_{E_03}(r_0)=0$.
As opposed to Ref.~\cite{Panta} this is not a perturbative
approach for small $\Nod$ but completely self-consistent.
In the adiabatic limit it consists of the solution of two nonlinear
equations. Apart from the step function approximation for the
medium mixing angle it is a more compact formulation of the
treatment in Ref.~\cite{Qian1}.

We can now define the effective adiabaticity coefficient for
mode $E_0$ within this self-consistent evolution as
\begin{equation}
  \Geff=\left.{V_{E_01}^2\over dV_{E_03}/dr}\right\vert_{r_0}=
  \left.{V_{E_01}^2\over\sqrt2\GF\left(d\Neff/dr\right)}
  \right\vert_{r_0}\,,\label{Geff}
\end{equation}
where $\Neff$ as a function of $r$ is given by Eq.~(\ref{Neff}).
This coefficient depends on $E_0$. For r-process nucleosynthesis
the most relevant modes are the high energy ones where
$F_{\nu_\mu}(E)>F_{\nu_e}(E)$. We therefore chose $E_0=30\MeV$
as a typical energy in the above equations.

Requiring $\Geff\geq3$
ensures that the phase dependent terms in Eq.~(\ref{phases}) are
suppressed by at least a factor 5 compared to the phase independent
terms and also allows a comparison with the bounds derived in
Ref.~\cite{Qian1}. For the model parameters given in section 3 this
condition leads to the excluded region in the
$\Delta-\sin^22\theta$ space to the right of the thick solid line shown
in Fig.~1A. Within a first order approximation one could
adopt the adiabatic limit for neutrinos with energies less than $E_0$
even for $\Geff\la3$ (for a more accurate approach in the
nonadiabatic case see Ref.~\cite{Qian1}) and average
over all phases in Eq.~(\ref{phases}). One could then extend the
bound down to $\Geff=0.23$ (corresponding to the critical
$\PLZ=0.7$ of Ref.~\cite{Qian}) which leads to the thin solid line in
Fig.~1A. In section 5 we will argue that it is indeed possible
to extend the analytical approach in this way.
The dashed and dotted lines in Fig.~1A correspond
to $\Geff=0.23$ and $\Geff=3$, respectively, neglecting neutrino
contributions to the effective potential. Their influence is
demonstrated by comparing respective curves to the same $\Geff$.
The dashed line corresponds to the
limit from Ref.~\cite{Qian}. The results are similar to the ones
obtained in Refs.~\cite{Qian,Qian1}.

It should be mentioned that Eqs.~(\ref{Nod}), (\ref{sinfin}) can have
multiple solutions, especially for small $\Delta$. This leads to
an uncertainty of typically a factor 2 to 3 in the bound on
$\sin^22\theta$ for $\Delta\la100\eV^2$. These bounds depend
of course also on the actual model adopted for the hot bubble phase.
As an illustration of this dependence Fig.~1B was calculated for
the same model
parameters as Fig.~1A apart from a neutrino luminosity increased
by a factor 2 which could well be within the uncertainty of actual
supernova models. In this case flavor conversion becomes less adiabatic mainly
because the self-consistent resonance position moves inward where
the electron density profile gets considerably steeper. This leads
to less stringent bounds on $\sin^22\theta$, an effect which is
largest for small $\Delta$. For $\Delta\sim1\eV^2$ the bound
is weakened by more than a factor 10. It should also be mentioned
that the turnover at low $\Delta$ of the bounds including the
neutrino induced potential turns out to be sensitive
to the density profile and could be uncertain within a factor
of about 3.

Finally, we note
that a similar analysis can be performed for
the shock reheating epoch by adopting a suitable density profile
for this earlier phase. As in Ref.~\cite{Qian1} it turns out that
neutrino self-interactions have a negligible influence on critical
mixing parameters derived in this situation. We will therefore
not consider this situation further here.

\section{A Numerical Model}
In order to test the analytical approach and extend the analysis
to the nonadiabatic case we have set up a
numerical model. As already seen in the analytical section
neutrino self-interactions play a negligible role during the
shock-reheating epoch. We will therefore restrict ourselves to
the r-process nucleosynthesis phase in this section.

To start with, for a fixed energy $E_0$ we define $r_0$ in this
section as the "zeroth order" resonance point determined by
neglecting self-interactions, i.e. as the solution of the equation
\begin{equation}
  \sqrt2\GF N_e={\Delta\over2E_0}\cos2\theta\,.\label{resc}
\end{equation}
Furthermore, we define a suitable variable $u=k(r-r_0)$ with
\begin{equation}
  k=\gamma^{-1}{\Delta\over2E_0}\sin2\theta=
  \left\vert N_e^\prime/N_e\right\vert_{r_0}\cot2\theta
  \label{kdef}
\end{equation}
(a prime denotes derivative with respect to $r$),
where the adiabaticity parameter $\gamma$ is given by
\begin{equation}
  \gamma={\Delta\sin^22\theta\over2E_0\cos2\theta}
  \left\vert N_e^\prime/N_e\right\vert_{r_0}^{-1}\,.\label{gamdef}
\end{equation}
The usual MSW resonance width corresponds to $\Delta u=2$ in
this coordinate. Next, we have to discretize Eq.~(\ref{evo2}) in
energy space. To this end we define central values
$E_k=E_0(1+k\tan2\theta/d)$ of equally wide energy bins
for $k=-n,\cdots,n$ where $d$ determines the bin width
$\Delta E=E_0\tan2\theta/d$. Since the energy resolution should at least be
of the order of the energy range over which neutrinos resonate within
one resonance length, $d$ should be not smaller than $\sim0.5$. As
we will show below
the results of the simulations do not depend sensitively on $d$ as
long as $d\ga0.5$. For the total polarization $\P$
appearing in Eq.~(\ref{evo2}) we write
\begin{equation}
  \P=(N_{\nu_e}+N_{\nu_\mu})\sum_i w_i\P_i\,,\label{Pdef}
\end{equation}
where the sum runs over all modes taken into account and
\begin{equation}
  w_i={\int_{E_i-\Delta E/2}^{E_i+\Delta E/2}dE
  \left[F_{\nu_e}(E)+F_{\nu_\mu}(E)\right]\over
  N_{\nu_e}+N_{\nu_\mu}}\,.\label{widef}
\end{equation}
Taking everything together and approximating $N_e(r)$ linearly at the
resonance Eq.~(\ref{evo2}) transforms into
\begin{equation}
  {d\over du}\P_i=\gamma\left[\left(\begin{array}{c}
  -E_0/E_i \\ 0 \\
  -u+\cot2\theta\left(1-E_0/E_i\right)
  \end{array}\right)+
  {g\over\sin2\theta}\sum_j w_j\P_j\right]\times\P_i
  \,,\label{numdef}
\end{equation}
where the self-coupling constant $g$ is defined by
\begin{equation}
  g=F(r_0)\left.{N_{\nu_e}+N_{\nu_\mu}\over N_e}
  \right\vert_{r_0}\,.\label{gdef}
\end{equation}
For notational simplicity antineutrinos were neglected in this
derivation. In principle one can include the antineutrino contribution
in an analytical way by substituting $N_e\to\Neff\equiv N_e+F(r)
\left(N_{{\bar\nu}_\mu}-N_{{\bar\nu}_e}\right)$ because antineutrinos
encounter no resonances and $\cos2{\bar\theta}_m\sim1$. Under the same
approximations and for not too high neutrino luminosities this leads
to only a moderate change in the parameters $\gamma$ and $g$ and leaves
the generic form of Eq.~(\ref{numdef}) unchanged. We also neglected
the radial dependence of the nonlinear term in Eq.~(\ref{numdef})
which can become important for $\vert u\vert\gg1/\sin2\theta$.
In the simulations below this can only influence the low energy
neutrinos which encounter resonances at highly negative $u$. However,
because low energy neutrinos in general tend to
be converted more adiabatically than high energy ones the
transition at high energies is quite insensitive to the situation
at highly negative $u$ as long as this transition is not too nonadiabatic.
The same remark holds for the sensitivity to deviations of the
electronic density slope $N_e(r)$ from the linear approximation
far from $u=0$.

The coupled multi mode nonlinear differential
equations (\ref{numdef}) have to be integrated on an interval
$[\umin,\umax]$ chosen such that all modes taken into account
encounter their resonances within this interval. As initial
conditions we choose flavor eigenstates,
\begin{equation}
  \P_i(\umin)={F_{\nu_e}(E_i)-F_{\nu_\mu}(E_i)\over
  F_{\nu_e}(E_i)+F_{\nu_\mu}(E_i)}\,{\bf e_3}\label{ini}
\end{equation}
(${\bf e_3}$ is the unit vector into the positive 3-direction).
For small vacuum mixing angles a suitable measure for the nonadiabatic
transition probability for high energy neutrinos is then given
by
\begin{equation}
  P_r={1\over2}\left(1+{\sum_{E_i\geq E_0}w_i\left[\P_i(\umax)
  \right]_3\over\sum_{E_i\geq E_0}w_i\left[\P_i(\umin)\right]_3}
  \right)\,.\label{Pr}
\end{equation}

For numerical simulations based on Eq.~(\ref{numdef})
$\sin^22\theta$, $\gamma$ and $g$ are the natural parameters. In
the actual problem for given $\sin^22\theta$ the concrete
values of $\gamma$ and $g$ in terms of $\Delta$ are given by
Eq.~(\ref{gamdef}) and Eq.~(\ref{gdef}), respectively. As in the
analytical section we have chosen $E_0=30\MeV$ but the results
do not change considerably for $E_0$ in the interval $[25\MeV,30\MeV]$. We
basically consider two cases in the following.

\subsection{Simulations in the Adiabatic Limit}
To become familiar with typical properties of solutions to
Eq.~(\ref{numdef}) we first choose parameters corresponding to
mixing near the boundary of the adiabatic regime. Here the
computational effort is less than in the nonadiabatic regime.
Adopting the model described in section 3 the mixing parameters
$\Delta=5\eV^2$ and $\sin^22\theta=10^{-3}$
lead to $\gamma\sim5$, $g\sim1$ at the resonance point. Its
location is determined by taking electrons and antineutrinos
(in the analytical way described above) into account. These parameters serve
as input values for Eq.~(\ref{numdef}). For the simulation shown
in Fig.~2A the energy interval spanned by the modes taken into
account is $[20\MeV,40\MeV]$. This simulation was actually performed
for both $d=1$ and $d=2$ resulting in $P_r=0.155$ and $P_r=0.148$ (Fig.~2A
shows only the $d=2$ case). This demonstrates insensitivity of
our results to $d$. Figs.~2B and 2C were therefore produced for
$d=1$. Fig.~2B is based on the same parameters as Fig.~2A except
that the self-coupling $g$ was enhanced by a factor 2. For the
simulation shown in Fig.~2C, which was done for exactly the same
parameters as Fig.~2B, more modes were taken into account corresponding
to an energy interval $[12.5\MeV,40\MeV]$. This led to a
considerably decreased $P_r=0.137$ compared to the case in Fig.~2B.

Figs.~2 demonstrate two common features which we observed within the
parameter range considered in our simulations: First,
increasing $g$ while keeping all other parameters fixed tends to
decrease the adiabaticity of flavor conversion. Second, taking
more low lying energy modes into account for fixed other
parameters tends to increase adiabaticity which depends roughly
logarithmically on the energy interval spanned by the modes
taken into account. This second feature is indeed what would be expected
from the analytical estimate [Eqs.~(\ref{Nod}) and (\ref{VE1})]
of the effective adiabaticity parameter Eq.~(\ref{Geff}): At
least as long as $\Nod\la\Neff$ (which is the case for the model
parameters presented in section 3) modes with
$F_{\nu_\mu}(E)>F_{\nu_e}(E)$, i.e. with $E\geq20\MeV$
tend to decrease $\Geff$ whereas low energy modes
tend to increase it. Because these low energy modes are considerably higher
occupied we expect to underestimate the adiabaticity for the energy
intervals chosen in our simulations. This also nicely demonstrates that
the numerical problem is {\it nonlocal} in energy space, i.e. in
principle all modes with considerable occupation have to be followed.
This has to do with the energy dependence of the medium mixing
angle which enters into the integral in Eq.~(\ref{Nod}).

In any case in the examples
to Figs.~2 $P_r$ is always considerably smaller than 0.5.
This confirms that $\Delta=5\eV^2$, $\sin^22\theta=10^{-3}$
can be excluded as already predicted by the conservative
analytical bounds (see thick solid line in Fig.~1A).

\subsection{Simulations in the Nonadiabatic Case}
The second mixing parameter combination we consider, $\Delta=10\eV^2$,
$\sin^22\theta=10^{-4}$ is not excluded by the conservative
analytical bound but is barely excluded by a naive extension of these bounds
(see thin solid line in Fig.~1A) and also by the bounds derived
in Ref.~\cite{Qian1}.
Within our numerical framework these mixing parameters correspond
to $\gamma\sim1$ and $g\sim1.5$. Based on our discussion of the
previous simulations we expect a simulation for $\gamma=1$, $g=2$
on an energy interval $[13\MeV,39\MeV]$ to underestimate adiabaticity.
The results are shown in Figs.~3. From Fig.~3A it can be seen
that in this case individual modes oscillate
a lot in a seemingly chaotic way due to the strong inter-mode
coupling. Nevertheless there is a continuous transformation
of high energy muon neutrinos into electron neutrinos until the
last mode has gone through its resonance. This is shown by
the evolution of the third component of the neutrino induced
effective potential $V_{\nu3}$ in Fig.~3B. Also shown in this
figure is a typical transverse component of the effective
potential $V_T$. Its high values during
conversion of neutrinos in low energy modes causes the conversion to be
much more adiabatic in these modes than in the higher energy
modes as already mentioned earlier. In any case since $P_r=0.500$
this simulation demonstrates that
$\Delta=10\eV^2$, $\sin^22\theta=10^{-4}$ still violates the condition
$\PLZ\ga0.7$ necessary for r-process nucleosynthesis to work in
supernovae~\cite{Qian}.

It also turns out that an analytical expression of the kind of
Eq.~(\ref{Geff}) still gives a rough estimate of $\Geff$ and
$P_r$ in these simulations. This indicates that analytical
estimates for flavor conversion efficiencies can still be used
to extend the mixing parameter bounds based on r-process
nucleosynthesis to the nonadiabatic regime. The resulting bounds depend
somewhat on the supernova model and are similar to the thin solid lines
in Figs.~1 or the bounds derived in Ref.~\cite{Qian1}.

Finally, it should be mentioned that runs performed with only
the flavor diagonal part of the effective potential taken into
account lead to substantially different behaved solutions and
transition efficiencies. This demonstrates nicely that the
off diagonal refractive index revealed by the quantum kinetic
approach has direct physical consequences.

\section{Conclusions}
We have analyzed oscillations among electron and muon or tau
neutrinos in the hot bubble region above the neutrinosphere
after supernova core bounce. We have shown that in the adiabatic
limit conservative bounds on the mixing parameters can be
derived from an analytical approach including neutrino
self-interaction contributions to the self-consistent effective
potential. These bounds concern neutrino masses between about $1\eV$
and $100\eV$ where neutrinos could serve as hot dark matter.
Because in the lower mass range these limits are one to two
orders of magnitude less restrictive than former limits which
neglected neutrino contributions we explored the parameter range
beyond these bounds by numerical simulations. That way it was
demonstrated that analytical estimates of flavor transition
probabilities can be extended within sufficient accuracy to the
nonadiabatic regime as was done in earlier work neglecting
phase effects. The resulting bounds are therefore similar
to the ones derived in Ref.~\cite{Qian1}. This conclusion is in
contrast to Ref.~\cite{Panta} where it was claimed that due
to the oscillation phases reliable constraints can not be
derived for mixing masses smaller than about $25\eV$.
We also commented on the dependence of the resulting bounds
on the actual supernova model adopted, especially the neutrino
luminosity and the density profile. Compared to former
limits on neutrino mixing based on calculations only
incorporating the electronic part to the effective potential
these bounds tend to be slightly weakened in the mixing angle
but at the same time extend to somewhat lower mixing masses.

\section*{Acknowledgments}
I am very grateful to Yong-Zhong Qian, Georg Raffelt and Evalyn Gates
for their comments on the manuscript. I also would like to thank
George Fuller for helpful
discussions. This work was supported by the DoE and by NASA through grant
NAGW-2381 at Fermilab and by the Alexander-von-Humboldt
Foundation.

\newpage

\section*{Figure Captions}
\bigskip

\noindent{\bf Figure 1A:}
Exclusion plot in the $\Delta-\sin^22\theta$ plane. The area
to the right of the thick solid line (corresponding to $\Geff=3$) is
excluded by the analytical approach of section 4. This is regarded
as a safe bound. The thin solid line corresponds to a naive
extension (see section 4) of the analytical bound down to
$\Geff=0.23$ ($\PLZ=0.7$). These lines are similar to
corresponding bounds in Ref.~\cite{Qian1}. The
dotted line corresponds to $\Geff=3$ without taking self-interaction
effects into account. For comparison with Ref.~\cite{Qian} the
dashed line corresponds to $\PLZ=0.7$ with the same parameters used
but again without self-interactions taken into account. The turnover
at low $\Delta$ of the solid lines is sensitive to the density
profile and could be uncertain by about a factor 3.

\medskip
\noindent{\bf Figure 1B:}
Same as Figure 1A but for a neutrino luminosity $L$ increased by a
factor 2. As discussed in section 4 this demonstrates the dependence of
the analytical bounds on the supernova model.

\medskip
\noindent{\bf Figure 2A:}
Numerical simulation of the model of section 5 with $\gamma=5$,
$g=1$, $\sin^22\theta=10^{-3}$ and $d=2$. 41 energy modes equidistant
on the interval $[20\MeV,40\MeV]$ were taken into account. The three
curves show the evolution of the third polarization component to the 20,
30 and $40\MeV$ energy mode, respectively, in direction of decreasing
initial value for $P_3$. The quantity $P_r$ defined in Eq.~(\ref{Pr})
measuring the level crossing probability at high energies
relevant for r-process nucleosynthesis is given by $P_r=0.148$.

\medskip
\noindent{\bf Figure 2B:}
Same as Figure 2A but for $d=1$ (21 modes) and a self-coupling
$g$ increased by a
factor 2 leading to an enhanced $P_r=0.301$. This demonstrates the
dependence of $\Geff$ on the supernova model.

\medskip
\noindent{\bf Figure 2C:}
Same as Figure 2B except that the oscillations were followed on
the larger energy interval $[12.5\MeV,40\MeV]$ corresponding to
29 modes. The curve to the lowest energy has a positive initial
value for $P_3$ because electron
neutrinos are more abundant at these energies. The result $P_r=0.137$ is
smaller compared to the previous case for reasons discussed in the text.

\medskip
\noindent{\bf Figure 3A:}
Numerical simulation of the model of section 5 with $\gamma=1$,
$g=2$, $\sin^22\theta=10^{-4}$ and $d=0.5$. 43 energy modes
equidistant on the interval $[13\MeV,39\MeV]$ were taken into
account. The three curves show the evolution of the
third polarization component to the 13,
30 and $39\MeV$ energy mode, respectively, in direction of decreasing
initial value for $P_3$. The quantity $P_r$ defined in Eq.~(\ref{Pr})
measuring the level crossing probability at high energies
relevant for r-process nucleosynthesis is given by $P_r=0.500$.

\medskip
\noindent{\bf Figure 3B:}
Evolution of the effective potential as defined in Eq.~(\ref{numdef})
for $E=30\MeV$ for the simulation on which Fig.~3A is based on. Shown are
the first component $V_1$ (dotted line), the modulus of
the tangential component $V_T$ (solid line) and the neutrino contribution
to the third component $V_{\nu3}$ (dashed line). Note that due
to Eq.~(\ref{numdef}) the resonance position of the mode to energy $E$
satisfies $u=\cot2\theta(1-E_0/E)+V_{\nu3}(u)$.


\begin{thebibliography}  {std}

\bibitem{Mikheyev} S.~P.~Mikheyev and A.~Yu.~Smirnov, {\sl Yad.
Fiz.} {\bf 42} (1985) 1441 [{\sl Sov. J. Nucl. Phys.} {\bf 42}
(1985) 913]; {\sl Nuovo Cimento} {\bf C 9} (1986) 17;
L.~Wolfenstein, {\sl Phys. Rev.} {\bf D 17} (1978) 2369.

\bibitem{Bethe} H.~A.~Bethe, {\sl Phys. Rev. Lett.} {\bf 56}
(1986) 1305.

\bibitem{Haxton} W.~C.~Haxton, {\sl Phys. Rev. Lett.} {\bf 57}
(1986) 1271.

\bibitem{Parke} S.~J.~Parke, {\sl Phys. Rev. Lett.} {\bf 57}
(1986) 1275; S.~J.~Parke and T.~P.~Walker, {\sl Phys. Rev. Lett.}
{\bf 57} (1986) 2322.

\bibitem{Kuo1} For a review see T.~K.~Kuo and J.~Pantaleone,
{\sl Rev. Mod. Phys.} {\bf 61} (1989) 937.

\bibitem{Fuller} G.~M.~Fuller, R.~Mayle, B.~S.~Meyer and
J.~R.~Wilson, {\sl Ap. J.} {\bf 389} (1992) 517.

\bibitem{Kuo2} T.~K.~Kuo and J.~Pantaleone, {\sl Phys. Rev.}
{\bf D 37} (1988) 298.

\bibitem{Notzold} D.~N\"otzold, {\sl Phys. Lett.} {\bf B 196}
(1987) 315.

\bibitem{Qian} Y-Z.~Qian, G.~M.~Fuller, G.~J.~Mathews,
R.~W.~Mayle, J.~R.~Wilson and S.~E.~Woosley, {\sl Phys. Rev.
Lett.} {\bf 71} (1993) 1965.

\bibitem{early} A.~Kostelecky, J.~Pantaleone and S.~Samuel, {\sl
Phys. Lett.} {\bf B 315} (1993) 46; A.~Kostelecky and S.~Samuel,
{\sl Phys. Lett.} {\bf B 318} (1993) 127; {\sl Phys. Rev.} {\bf
D 49} (1994) 1740; S.~Samuel, {\sl Phys. Rev.} {\bf D 48} (1993)
1462.

\bibitem{Panta} J.~Pantaleone, Report IUHET-276 (1994).

\bibitem{Qian1} Y-Z.~Qian and G.~M.~Fuller, Univ. of Washington
preprint DOE/ER/40561-150-INT94-00-63 (1994).

\bibitem{Cline} J.~M.~Cline, Preprint UMN-TH-1302-94, TPI-MINN-94/23-T
(1994).

\bibitem{Sigl} G.~Sigl and G.~Raffelt, {\sl Nucl. Phys.} {\bf
B406} (1993) 423.

\bibitem{RS} G.~Raffelt and G.~Sigl, {\sl Astropart. Phys.} {\bf
1} (1993) 165.

\bibitem{quakin} similar equations can be found in:
A.~D.~Dolgov, {\sl Yad. Fiz.} {\bf 33} (1981) 1309 [{\sl Sov. J.
Nucl. Phys.} {\bf 33} (1981) 700]; M.~A.~Rudzsky, {\sl
Astrophysics and Space Science} {\bf 165} (1990) 65;
B.~H.~J.~McKellar and M.~J.~Thomson, {\sl Phys. Lett.} {\bf B
259} (1991) 113; {\sl Phys. Rev.} {\bf D 49} (1994) 2710.

\bibitem{Burrows} A.~Burrows, {\sl Ann. Rev. Nucl. Part. Sci.}
{\bf 40} (1990) 181.

\end{thebibliography}
\end{document}